\documentclass[11pt,twoside]{article}
\usepackage[pdftex]{graphicx}
\usepackage{amsmath}
\usepackage{amssymb}
\usepackage{cite}
\usepackage{lineno}

\begin{document}
\newcommand{\pst}{\hspace*{1.5em}}

\newcommand{\rigmark}{\em Journal of Russian Laser Research}
\newcommand{\lemark}{\em Volume 30, Number 5, 2009}

\newcommand{\be}{\begin{equation}}
\newcommand{\ee}{\end{equation}}
\newcommand{\bm}{\boldmath}
\newcommand{\ds}{\displaystyle}
\newcommand{\bea}{\begin{eqnarray}}
\newcommand{\eea}{\end{eqnarray}}
\newcommand{\ba}{\begin{array}}
\newcommand{\ea}{\end{array}}
\newcommand{\arcsinh}{\mathop{\rm arcsinh}\nolimits}
\newcommand{\arctanh}{\mathop{\rm arctanh}\nolimits}
\newcommand{\bc}{\begin{center}}
\newcommand{\ec}{\end{center}}

\thispagestyle{plain}

\label{sh}


\begin{center} {\Large \bf
\begin{tabular}{c}
TUNABLE ABSORPTION IN NEAR INFRARED 
\\[-1mm]
BASED ON INDIUM TIN OXIDE

\end{tabular}
 } \end{center}

\bigskip

\bigskip

\begin{center} {\bf
Shuaishuai Hou$^{1}$, Lin Cheng$^{1,*,\dagger}$, Nan Meng$^{3}$, NaiXin Liu$^{2,3,4}$,  JianChang Yan$^{2,3}$, JinMin Li$^{1,2,3}$, RuiXin Yang$^{1}$and Kun Huang$^{1}$
}\end{center}

\medskip

\begin{center}
{\it
$^1$State Key Laboratory of Dynamic Testing Technology, North University of China, \\Taiyuan, China, 030051

\smallskip

$^2$Center of Materials Science and Optoelectronics Engineering, University of Chinese Academy of Sciences, \\
Beijing 100049, China

\smallskip
$^3$Research and Development Center for Wide Bandgap Semiconductors, Institute of Semiconductors, Chinese Academy of Sciences,\\ Beijing 100083, China
\smallskip

$^4$State Key Laboratory of Dynamic Measurement Technology, North University of China, \\Taiyuan 030006, China
\smallskip

$^*$Corresponding author: e-mail:kiki.cheng@nuc.edu.cn\\$\dagger$ These authors contributed equally to this work.}
\end{center}

\begin{abstract}\noindent
We propose a serrated ultra-broadband infrared absorber based on the multi-layer repetitive stacking of indium tin oxide (ITO) and silicon materials. ITO exhibits remarkable nonlinear optical effects under strong light, enabling rapid modulation of the light intensity. This absorber achieves continuously tunable absorption within a wide infrared light range by increasing the incident intensity. Though changing the size and materials of the absorber, its average absorption rate can reach a relatively stable level. Simulation results reveal wavelength-dependent electric field localization: shorter wavelengths concentrate in narrower upper regions, while longer wavelengths localize in wider lower areas. This design offers potential applications in near-infrared detectors, optical communication and solar energy.
\end{abstract}

\medskip

\begin{figure}[h!]
\centering\includegraphics[width=13cm]{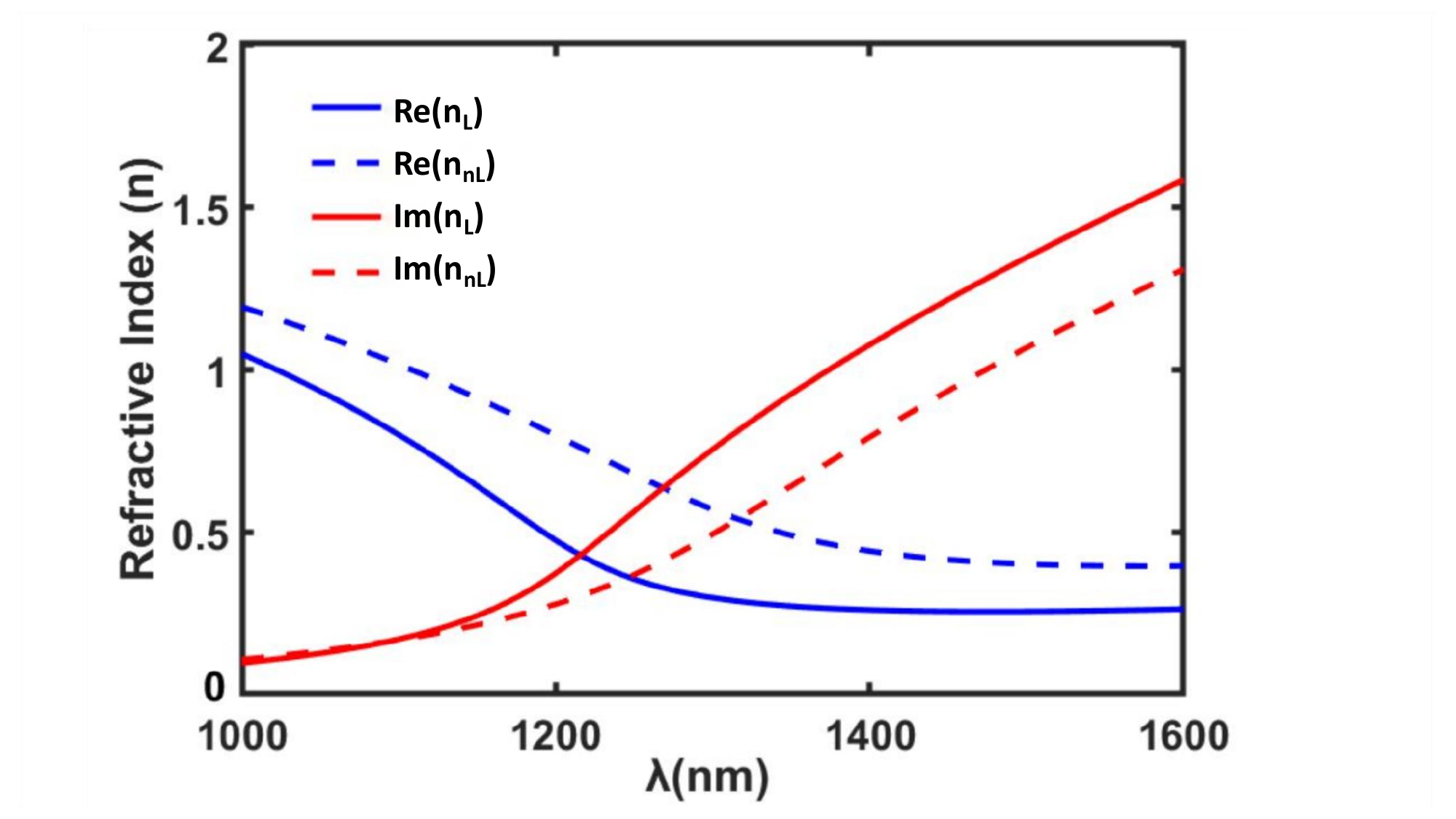}
\caption{Figure of the refractive index of ITO varying with wavelength~\cite{8}.}
\label{Fig.1}
\end{figure}

\pst
Indium tin oxide (ITO) is a transparent conductive material that combines excellent electrical conductivity and high transparency~\cite{1}. When the light intensity is below 100 GW/cm², the perturbation of the light field on electrons is weak, and a linear response occurs at this time, which is suitable for passive devices such as transparent conductive films\cite{2}. When the light intensity exceeds 100 GW/cm², intense light excitation leads to a sharp increase in the electron temperature, and the non-parabolicity of the conduction band is significantly manifested (as shown in Figure \ref{Fig.1})\cite{2}. This ultrafast (in the range of hundreds of femtoseconds) dynamic regulation ability makes ITO an ideal material for active photonic devices such as all-optical switches and ultrafast modulators, providing a new approach for optical communication and photonic computing. With the deepening of research, ITO materials are expected to exhibit unique advantages in more new technology fields and promote the innovation and progress of related technologies~\cite{3}~\cite{4}~\cite{5}.

Traditional infrared devices regulate electromagnetic waves by taking advantage of the inherent properties of natural materials. However, their ability to manipulate electromagnetic waves is limited, with low absorption rates and narrow operating bands~\cite{6}~\cite{7}. On the other hand, Metamaterials are artificially designed materials that possess special physical properties that are not found in natural
materials~\cite{8}. Through the artificial design of their structures and compositions, extraordinary characteristics, such as unique electromagnetic properties, can be achieved. The design of absorbers using metamaterials has brought about new ideas and methods to solve related problems~\cite{9}.
However, at present, absorbers operating in the near-infrared band still have some deficiencies. Specifically, the absorption bands are relatively narrow, and it is difficult to effectively regulate multiple structural parameters~\cite{10}. In response to these existing shortcomings, this paper designs a multilayer structure based on indium tin oxide (ITO) and silicon materials~\cite{11}. Through a series of structural optimization works, the absorber achieves high absorption rate regulation in a relatively wide band near-infrared range. In applications such as optical communication and infrared detection, a broadband infrared absorber with precise controllability is crucial~\cite{12}, as it can greatly improve the performance and signal processing capabilities of the system~\cite{13}. To achieve more precise regulation of absorption performance, we have attempted various methods to adjust the structure and dimensions of the materials. By utilizing the unique optical coupling effect and light field localization phenomenon in the zigzag structure, we have designed a near-infrared absorber that exhibits excellent tunable and continuous absorption characteristics in the infrared range from 1000 to 1600 nm. 
\begin{figure}[htb!]
\centering\includegraphics[width=8cm]{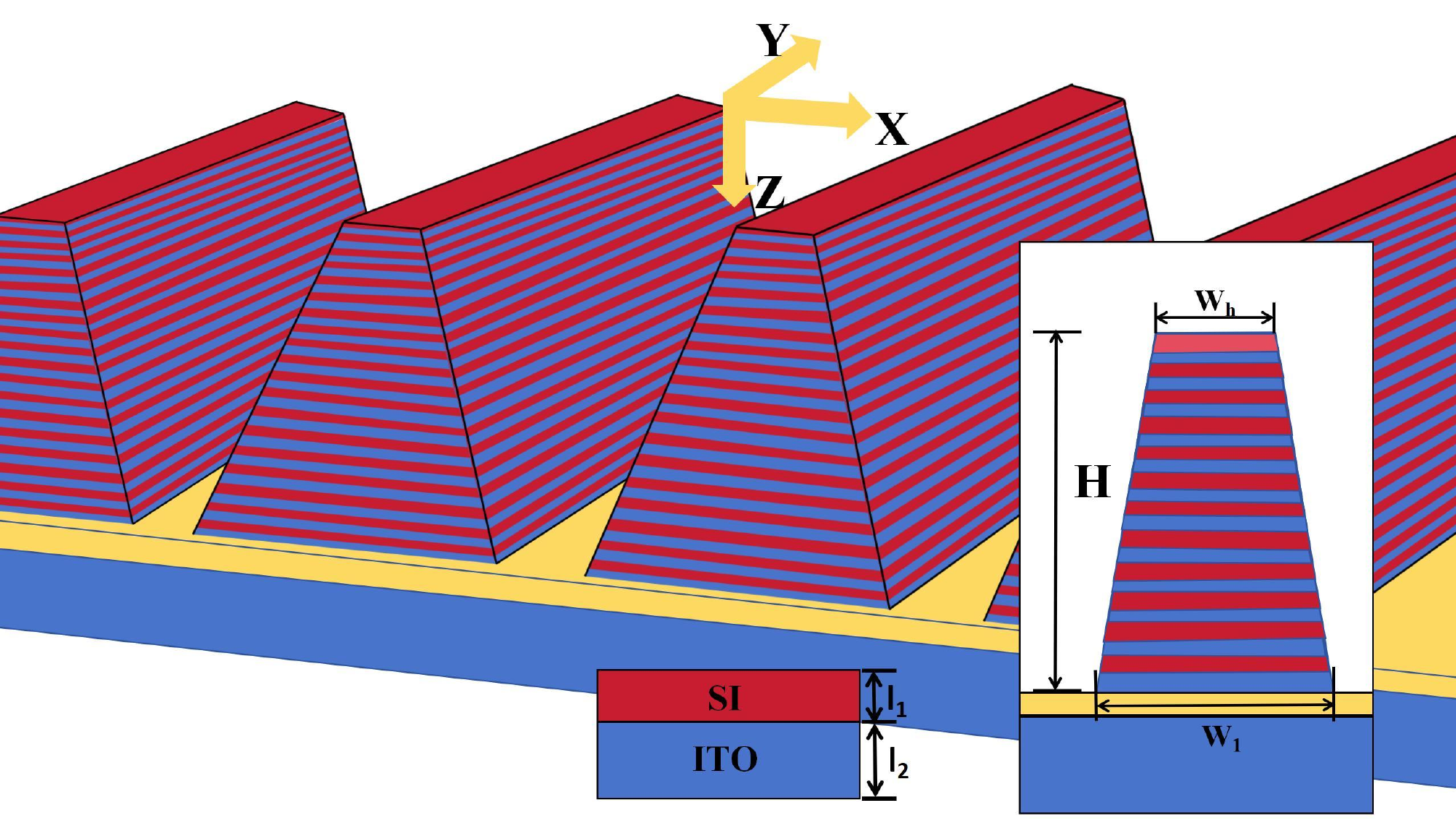}
\caption{Schematic diagram of the zigzag - shaped metamaterial thin-film absorber.$H = 1000$ nm, $W_h = 150$ nm, $W_l = 600$ nm, $l_1 = 120$ nm, $l_2 = 40$ nm.}
\label{Fig.2}
\end{figure}
At present, research on large-scale regulation in near-infrared light using linear and nonlinear materials is relatively scarce~\cite{14}. We have achieved tunable research by designing a serrated absorber with appropriate dimensions~\cite{15}. Compared with the spiky cross absorption curve obtained from a near-infrared absorber with only one set of Si (assuming Si is a specific material) and ITO superposition, our 11-set serrated infrared absorbers exhibit smooth and regular tunable absorption curves within a wide range of infrared bands~\cite{16}. Specifically, the zigzag-shaped material employed in this study consists of alternating layers of indium tin oxide (ITO) and Si (see Figure \ref{Fig.2}). The thickness of the ITO layer is set at 120 nm, while that of the silicon layer is 40 nm. 
The plate is fabricated into a zigzag shape with a periodic interval. The period of the zigzag structure is 800 nm, the top width of each zigzag is 150 nm, and the bottom width is 600 nm. To effectively block all transmissions, a gold film with a sufficiently large thickness is added beneath the zigzag-shaped plate~\cite{17}.

\begin{figure}[ht!]
\centering\includegraphics[width=10cm]{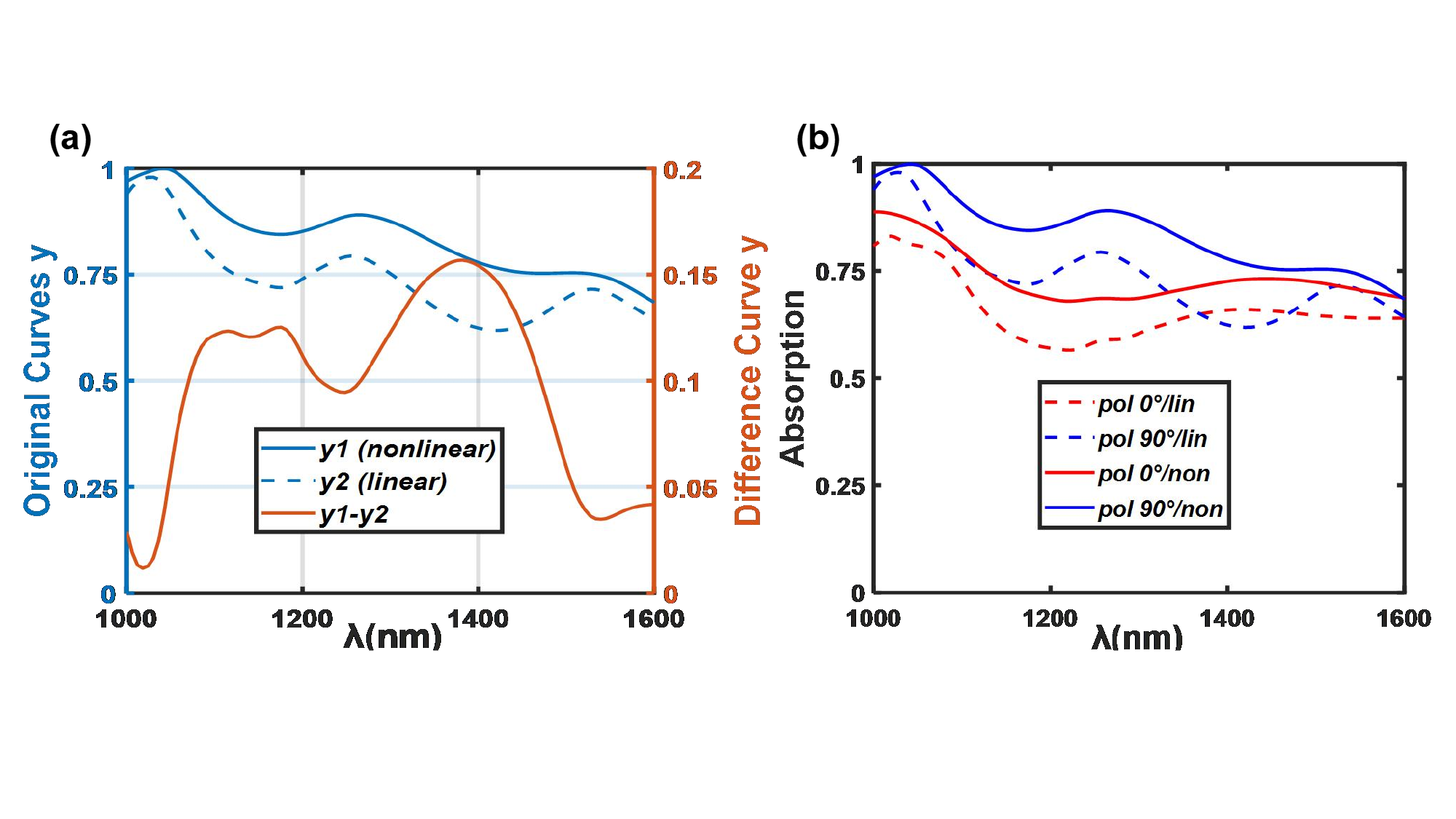}          
\caption{(a) Absorptance and tunable range of the zigzag absorber under illumination in the wavelength range of 1000 to 1600 nm. There are 11 pairs of ITO and silicon. (b) Absorptance at different polarization angles.}
\label{Fig.3}
\end{figure}

By simulating the impact of a plane wave with TE - TM-polarized light incident along the z - direction on the structure, the absorption rate (defined as \(A = 1 - R - T\), where \(T = 0\)) is calculated based on Poynting's theorem. The absorption spectra obtained under normal incidence (Figure \ref{Fig.2}a, \(y_1,y_2\)) indicate excellent absorption performance. The absorptivity is higher than \(85\%\) in the range of \(1000\) to \(1350\) nm, with an average absorptivity of \(86\%\), and the maximum adjustable range reaches \(0.157\) (Figure \ref{Fig.3}a, \(y_1 - y_2\)). Moreover, when the incident light remains unchanged, changing the polarization angle will result in different absorption efficiencies. From the simulation results of the \(0^{\circ}\) and \(90^{\circ}\) polarization angles (Figure \ref{Fig.3}b), we can observe that compared with the \(0^{\circ}\) polarization angle, the results obtained at the \(90^{\circ}\) polarization angle are more ideal. There is a significantly improved absorptivity, as well as better regulation performance for both linear and non-linear characteristics.

To evaluate the overall light conversion ability of the absorber in light-collecting and optical detector applications, it is also meaningful to calculate the total absorbed energy efficiency, which is defined as the integral of the ratio of the absorbed energy to the total incident energy over the considered energy band, that is,

\begin{equation}
 \Delta = \frac{\int_{\omega_1}^{\omega_2} \eta(\omega) \mathrm{d}\omega}{\omega_2 - \omega_1}.
\end{equation}
Here, the energy band corresponding to the considered wavelength range from 1000 to 1600 nm is taken into account. The value of \(\Delta\) for our near-infrared absorber is 85.9, which indicates that a large portion of the incident energy within the considered range can be absorbed. 

\begin{figure}[ht!]
\centering\includegraphics[width=10cm]{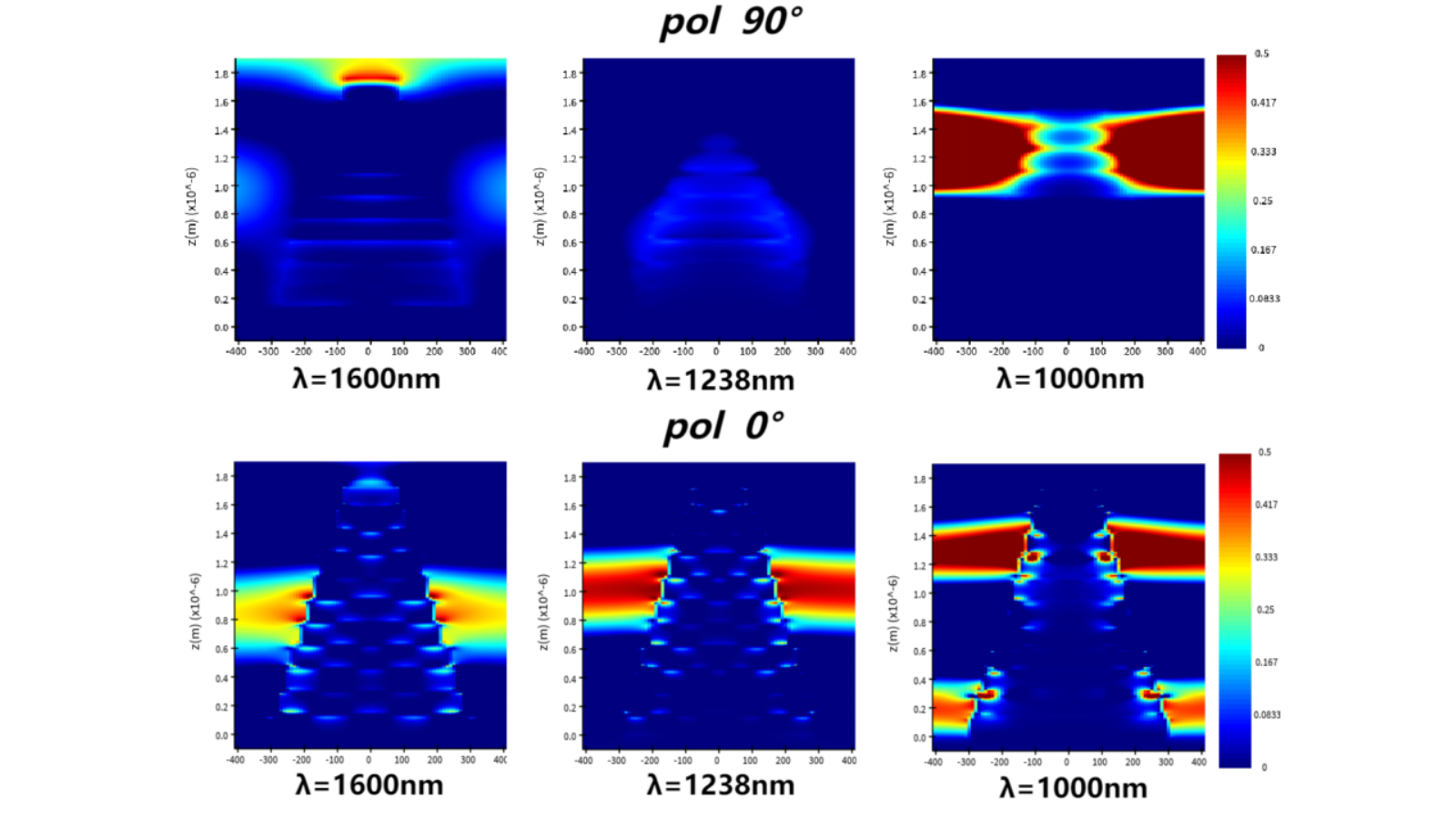}
\caption{Electric field diagrams of the absorption of light sources with different wavelengths by the zigzag structure at polarization angles of $0^\circ$ and $90^\circ$. }
\label{Fig.4}
\end{figure}

To delve into the light absorption mechanism of the zigzag absorber, a comprehensive study was conducted on the electromagnetic field distribution of the zigzag structure at three distinct wavelengths ($\lambda_0 = 1195$ nm, $\lambda_1 = 1253$ nm, and $\lambda_2 = 1507$ nm). The corresponding results are presented in the color-coded map in Figure \ref{Fig.4}.
The findings of this investigation indicate that at a polarization angle of 90°, light of varying wavelengths tends to localize at different regions of the zigzag absorber. Specifically, for light with a relatively short wavelength ($\lambda_0 = 1195$ nm), the electric field is predominantly concentrated within the upper-layer structure, suggesting the occurrence of a pronounced surface plasmon resonance. When the wavelength approaches $\lambda_1 = 1253$ nm, the light in the absorption band's middle accumulates at the zigzag configuration's midsection. In the case of incident light with a longer wavelength ($\lambda_2 = 1507$ nm), the energy is localized at the broader serrations in the lower part of the zigzag plate, with scarcely any significant electric field concentration observed around the top.
Conversely, at a polarization angle of 0°, the electric field profiles of the zigzag absorber for different wavelengths remain largely unchanged. This observation corroborates the fact that the absorptance at a polarization angle of $90^\circ$ surpasses that at a polarization angle of $0^\circ$. Moreover, the analysis of the electric field within the wavelength range of 1000 nm to 1600 nm validates the absorption mechanism of the zigzag structure.

The unique zigzag structure plays a crucial role~\cite{18}. It is obtained by etching the material into a zigzag shape using special micro- and nano-fabrication techniques~\cite{19}. This structure can not only serve as a complex optical-field coupling platform~\cite{20}. By changing the shape parameters of the material's zigzags (such as the size, angle, and spacing of the zigzags), it can achieve the coupling and absorption of infrared light at different wavelengths, thereby realizing the tunability of the absorption range~\cite{21}. Moreover, due to the gradual change in the optical properties at the material interface, it helps to improve the absorption efficiency and the continuity of the absorption range~\cite{22}~\cite{23}.
Meanwhile, the multi-layer stacked structure also provides conditions for multiple reflections and absorptions of light within it, further enhancing the overall absorption effect. It offers a promising absorber solution for applications in fields such as infrared light detection, thermal imaging, and optical communication. By deeply understanding and utilizing the physical mechanisms within this structure, it is expected to further expand its application prospects in infrared light applications with a wider wavelength range and higher performance requirements, thus promoting the progress and development of related technologies~\cite{24}.

In summary, we have designed a metamaterial-based ultra-broadband near-infrared tunable absorber. After the metamaterial is processed into a zig-zag shape, it exhibits a high absorption rate, a wide operating band, and strong tunability~\cite{24}. By designing the shape parameters of the zigzags, we can achieve the coupling and absorption of near-infrared light at different wavelengths, thereby realizing the tunability of the absorption range. Moreover, due to the gradual change in the optical properties at the material interface, the absorption efficiency and the continuity of the absorption range are improved.
Through calculations, the value of $\Delta$ for our near-infrared absorber is 85.9, indicating that a large portion of the incident energy within the considered range can be absorbed~\cite{25}. Analyses of the electric- and magnetic-field maps at different wavelength bands also confirm the feasibility of the absorption mechanism of our zigzag absorber.
Our research is expected to be applied to the design of high-performance photovoltaic devices and thermal emitters~\cite{17}. It should be emphasized that our proposal is scalable and there is still room for further research~\cite{26}. Through further optimization of the structure and the utilization of algorithms such as machine learning and deep learning, better absorption performance and tunability can be achieved. 

\textbf{Ethical Approval}:not applicable

\textbf{Confilicts of Interest}: The authors declare no confilct of interest.

\textbf{Declarations}
 The statements, opinions and data contained in all publications are solely those of the individual
author(s) and contributor(s) and not of MDPI and/or the editor(s). MDPI and/or the editor(s) disclaim responsibility for any injury to
people or property resulting from any ideas, methods, instructions, or products referred to in the content.

\section{Data Availability Statement }
Data is available on request from the authors.
\section{Funding}
We would like to acknowledge the support of the National Natural Science Foundation of China (62305312), the Shanxi Province Natural Science Foundation,ina (202203021222021), the Research Project supported by the Shanxi Scholarship Council of China (2312700048MZ), and the Doctoral Science Foundation Foundation fellowship(2022M722923). Shanxi Provincial Teaching Reform and Innovation Project (2024YB008).
\section{Acknowledgement}
{Lin Cheng thanks Rasoul Alaee for his help. }


\end{document}